\documentclass[showpacs]{revtex4}
\usepackage{amsmath}
\usepackage{amssymb}
\usepackage{graphicx}

\begin{document}

\title{
Dipole magnetic field of neutron stars in $f(R)$ gravity
}
\author{{\bf Elizat Bakirova$^{1}$}}
\author{{\bf Vladimir~Folomeev$^{2}$}}
\email[{\it Email:}]{vfolomeev@mail.ru}
\affiliation{
$^1$Department of Theoretical Physics,
Kyrgyz National University, Bishkek, 720033, Kyrgyzstan\\
$^2$Institute of Physicotechnical Problems and Material Science of the NAS
of the
Kyrgyz Republic, 265 a, Chui Street, Bishkek, 720071,  Kyrgyzstan\\
}

\begin{abstract}
The structure of an interior dipole magnetic field of neutron stars in $f(R)$ gravity is considered.
For this purpose, the perturbative approaches are used when both the deviations from general relativity and the deformations
of spherically symmetric configurations associated with the presence of the magnetic field are assumed to be small.
Solutions are constructed which describe
relativistic, spherically symmetric
configurations consisting of a gravitating magnetized perfect fluid modeled by a realistic equation of state.
Comparing configurations from general relativity and modified gravity,
we reveal possible differences in the structure of the magnetic field
which occur in considering neutron stars in modified gravity.
\end{abstract}


\maketitle

\vspace{-.5cm}

\section{Introduction}

Neutron stars represent good objects for studying physical effects of strong gravity and
properties of matter under extreme conditions which are far outside
the realm of possible laboratory physics.
The mean density of neutron stars is of the order of nuclear density.
Such matter cannot be obtained in a laboratory, and, therefore, it is possible to describe it only by using
some theoretical models. Their verification is carried out by analyzing and interpreting the results of astronomical observations
with subsequent refinement of original theoretical models~\cite{Potekhin:2011xe}.

Usually neutron stars are studied in Einstein's general relativity (GR).
However, taking into account the presence of strong gravitational fields in such configurations,
strong-field effects resulting from a consideration of modified
gravity theories (MGTs) extending GR may in general begin to manifest themselves.
One of the simplest ways to modify GR is to change the Einstein gravitational Lagrangian $\sim R$ by the
modified Lagrangian $\sim f(R)$, where $f(R)$ is some function of the scalar curvature $R$.
Such  MGTs have initially been applied for the description of
the evolution of the very early Universe (see, e.g., the pioneering works~\cite{MTG_early}).
However, after the discovery of the accelerated expansion of the present Universe,
MGTs have found some interesting applications in modeling this acceleration. In particular,
instead of introducing a special substance, called dark energy, as it is done in general relativity,
in  MGTs such an acceleration can be provided without using any new fundamental ingredients (for a general review on the subject, see, e.g., Ref.~\cite{Nojiri:2010wj}).

At the present time MGTs are applied not only in describing the present acceleration  of the Universe but also, for example,
in considering thick brane models~\cite{thick_b}
and in modeling the rotation curves of galaxies~\cite{Capozziello:2004us}.
When scales of objects under investigation decrease further,
one may consider compact astrophysical configurations.
In particular, within the framework of $f(R)$ gravity,
this can be relativistic stars \cite{rel_star_f_R}
or such exotic objects as wormholes~\cite{wh_f_R}. However, effects of MGTs may also manifest themselves in considering less
exotic objects like neutron stars~\cite{Cooney:2009rr,Arapoglu:2010rz,Orellana:2013gn,Alavirad:2013paa,Astashenok:2013vza}.
Considering such objects within the framework of
different types of
$f(R)$ gravities and using  various equations of state for neutron matter, one can
reveal the allowed forms of $f(R)$ satisfying the observational constraints.

Another important ingredient of neutron stars is a strong magnetic field.
The measured surface magnetic fields of such objects are in the range of $10^{11}-10^{13}$~G (for the so-called ``classical pulsars'')
and can reach values of the order of $10^{15}$~G for magnetars.
Stellar magnetic fields are currently under intensive investigation
(see, e.g., the book~\cite{Mest}  on stellar magnetism in general and some works~\cite{Bocquet:1995je,magn_stars}
on particular questions
of the structure of strongly magnetized configurations, and also references therein).
Such studies are usually performed within the framework of GR. In turn, there are recent researches devoted to
a consideration of magnetic fields of compact stars in modified gravities~\cite{Hakimov:2013zoa}
and to study  the influence of superstrong magnetic fields ($\sim 10^{17}-10^{18} \text{G}$)
on the structure of neutron stars in MGTs~\cite{Cheoun:2013tsa,Astashenok:2014gda}.
The presence of such strong fields results in modifications of an
equation of state of neutron star matter and in changes in mass-radius relations.

In modeling magnetized neutron stars in modified gravity in Refs.~\cite{Cheoun:2013tsa,Astashenok:2014gda}, the approach is employed where
the distribution of a spherically symmetric magnetic field is given by some function of baryon density.
Here we want to consider the case where a
magnetic field is modeled in the form of an axially symmetric dipole field, whose distribution
is self-consistently determined from a solution of the Einstein-Maxwell equations.
In doing so, we assume that the energy density of the magnetic field
is much smaller than the energy density of the neutron matter. This permits us to make
a perturbative expansion of the field equations, where axisymmetric deformations of the configuration associated with the presence
of the magnetic field are regarded as second-order perturbations to a background spherically symmetric configuration~\cite{Konno:1999zv}.

In turn,  background spherically symmetric equilibrium solutions will be sought within the framework of the perturbative approach to $f(R)$ gravity
from Ref.~\cite{Cooney:2009rr}.
Then the resulting field equations are  second-order differential equations
with respect to metric functions, instead of fourth-order equations obtained in $f(R)$ gravity in general.

Working within the framework of the above perturbative approaches,  our goal will be to clarify
the influence which the effects of modified gravity have on the structure of the interior dipole magnetic field.

The paper is organized as follows:  In Sec.~\ref{statem_prob}  we present
the statement of the problem and derive the corresponding
equations in the MGT for the systems under consideration.
In Sec.~\ref{num_calc}  we numerically solve these equations
for two special choices of the function $f(R)$, for which we obtain distributions of the magnetic field
in GR and in the MGT.
Comparing the results,
we demonstrate the influence which the effects
of modified gravity have on the structure and strength of the magnetic field.
Also, in Sec.~\ref{ellipticity} we calculate the ellipticity of the configurations under consideration
appearing due to the presence of the axisymmetric magnetic field.
Finally, in
Sec.~\ref{conclusion} we summarize the results obtained.

\section{Statement of the problem and equations}
\label{statem_prob}

The purpose of the paper is to study the structure of the interior magnetic field of neutron stars in $f(R)$ gravity.
In doing so, we make use of the following simplifying assumptions~\cite{Konno:1999zv}:

\begin{itemize}
  \item
 We do not take into account rotational deformations
and consider only static equilibrium magnetized configurations consisting of neutron matter modeled by some realistic equation of state (EoS).
  \item
 The magnetic field is modeled in the form of an
axisymmetric, poloidal magnetic field produced by toroidal electric currents.
The presence of such a field results in a deviation of the shape of the configuration from spherical symmetry.
  \item
We consider the case where
 the magnetic field strength is of the order of
$10^{12}-10^{15}$~G. In this case
deviations from the spherical shape are negligible,
since the energy of the magnetic field is much smaller
than the gravitational energy~\cite{Sotani:2006at}.
This permits us to neglect in the zeroth approximation the deformations of the configuration associated
with the magnetic field and to consider such deformations as a second-order effect [see Eq.~\eqref{pert_metr} below].
\item The interior of a star is assumed to be
a perfectly conducting medium free of electric charges and fields.
\end{itemize}

Taking all this into account,
in Sec.~\ref{mod_eqs_gen} we write down the
background equations for a spherically symmetric case, and
in Sec.~\ref{mag_f_eqs} we derive  equations for the magnetic field.

\subsection{Background equations}
\label{mod_eqs_gen}

Here we consider modified gravity
with the action [the metric signature is $(+,-,-,-)$]
\begin{equation}
\label{action_mod}
S=-\frac{c^4}{16\pi G}\int d^4 x \sqrt{-g} f(R) +S_m,
\end{equation}
where $f(R)$ is an arbitrary nonlinear function of $R$ and $S_m$ denotes the action
of matter. Such a modification of Einstein's GR
gives new possibilities both in considering cosmological problems and in studying compact objects (see the Introduction).
However, when one varies this action, the resulting equations are in general  fourth-order
differential equations for metric functions, whose study encounters the known difficulties
(see, e.g., Refs.~\cite{DeDeo:2007yn,Cooney:2009rr}).

An alternative way is to introduce corrections to general relativity perturbatively, as suggested in Ref.~\cite{Cooney:2009rr}.
Instead of considering an exact theory in which the presence of higher-order derivatives
assumes the occurrence of new degrees of freedom, in this approach deviations from GR are assumed to be small.
Such an approach removes the extra degrees of freedom and strongly simplifies the theory.

Technically this approach assumes that
deviations from GR can be parameterized  by a single parameter $\alpha$. In this case
the gravitational Lagrangian density in \eqref{action_mod} is chosen in the form
\begin{equation}
\label{Lagr_mod_pert}
f=R+\alpha h(R)+ \mathcal{O}(\alpha^2),
\end{equation}
where $h(R)$ is an arbitrary function of $R$
 and $\mathcal{O}(\alpha^2)$ denotes the possible higher-order corrections in $\alpha$.
In turn, metric functions can be expanded in $\alpha$ as
$$
g_{ik}=g_{ik}^{(0)}+\alpha g_{ik}^{(1)}+\mathcal{O}(\alpha^2),
$$
where the superscript (0) means that the metric function is obtained in Einstein gravity, i.e., by solving
second-order differential equations.
Then  modified gravitational equations are obtained by varying the action \eqref{action_mod} with \eqref{Lagr_mod_pert}
with respect to the metric in the form
\begin{equation}
\label{mod_Ein_eqs_gen}
\left(1+\alpha h_R\right) G_i^k-\frac{1}{2}\alpha\left(h-h_R R\right)\delta_i^k+
\alpha \left(\delta_i^k g^{m n}-\delta_i^m g^{k n}\right)\left(h_R\right)_{;m;n}+\mathcal{O}(\alpha^2)=\frac{8\pi G}{c^4}T_i^k.
\end{equation}
Here $G_i^k\equiv R_i^k-\frac{1}{2}\delta_i^k R$ is the Einstein tensor, $h_R\equiv dh/dR$, and $;$ denotes the covariant derivative.

As a matter source in these equations, we take a perfect fluid with the energy-momentum
tensor
\begin{equation}
\label{fluid_emt}
T_i^k=\left(\varepsilon +p\right)u^k u_i-\delta_i^k p,
\end{equation}
where $\varepsilon$ and $p$ are the fluid energy density and pressure, respectively.

To derive the modified Einstein equations and the Tolman-Oppenheimer-Volkoff equation for the fluid, we choose
the spherically symmetric line element in the form
\begin{equation}
\label{metric_schw}
ds^2=e^{\nu}(dx^0)^2-e^{\lambda}dr^2-r^2 \left(d\Theta^2+\sin^2\Theta\, d\phi^2\right),
\end{equation}
where $\nu$ and $\lambda$ are functions of the radial coordinate $r$ only,
and $x^0=c\, t$ is the time coordinate.

Following Refs.~\cite{Cooney:2009rr,Arapoglu:2010rz}, let us look for perturbed solutions of the field equations by expanding the required functions in
$\alpha$ as follows:
$$
\nu=\nu_{(0)}+\alpha\nu_{(1)}+ \ldots, \quad \lambda=\lambda_{(0)}+\alpha\lambda_{(1)}+ \ldots, \quad
\varepsilon=\varepsilon_{(0)}+\alpha\varepsilon_{(1)}+ \ldots, \quad
p=p_{(0)}+\alpha p_{(1)}+ \ldots,
$$
where the subscript (0) again means that the functions are obtained from a solution of the Einstein equations.

Using this perturbative approach, one can derive the following $(^t_t)$ and $(^r_r)$ components of Eqs.~\eqref{mod_Ein_eqs_gen}:
\begin{eqnarray}
\label{mod_Einstein-00_gen}
&&\left(1+\alpha h_R\right)
\left[-e^{-\lambda}\left(\frac{1}{r^2}-\frac{\lambda^\prime}{r}\right)+\frac{1}{r^2}\right]-
\alpha\left\{\frac{1}{2}\left(h-h_R R\right)+e^{-\lambda}\left[h_R^{\prime\prime}-\left(\frac{1}{2}\lambda^\prime-\frac{2}{r}\right)h_R^{\prime}\right]
\right\}
=\frac{8\pi G}{c^4} \varepsilon,
 \\
\label{mod_Einstein-11_gen}
&&\left(1+\alpha h_R\right)
\left[-e^{-\lambda}\left(\frac{1}{r^2}+\frac{\nu^\prime}{r}\right)+\frac{1}{r^2}\right]-
\alpha\left[\frac{1}{2}\left(h-h_R R\right)+e^{-\lambda}\left(\frac{1}{2}\nu^\prime+\frac{2}{r}\right)h_R^{\prime}
\right]
=-\frac{8\pi G}{c^4} p,
\end{eqnarray}
where the prime denotes differentiation with respect to $r$ and the right-hand sides have been taken from \eqref{fluid_emt}.
Assuming that the external spacetime is  described by the Schwarzschild solution,
one can introduce a new function $M(r)$, defined as
\begin{equation}
\label{metr_g11}
e^{-\lambda}=1-\frac{2 G M(r)}{c^2 r}.
\end{equation}
Then the surface value of $M$ can be regarded as a gravitational mass of the configuration.
Let us expand, as before, $M$ in $\alpha$ as
$M=M_{(0)}+\alpha M_{(1)}$, where
$$
M_{(0)}=\frac{4\pi}{c^2}\int_0^{r_b} \varepsilon_{(0)} r^2 dr
$$
is the general relativistic mass of a fluid sphere with the boundary located at the
radius $r_b$.
Substituting now \eqref{metr_g11} into \eqref{mod_Einstein-00_gen}, we have
\begin{equation}
\label{mass_eq}
\frac{d M}{d r}=\frac{4\pi}{c^2} r^2 \varepsilon-\alpha \frac{c^2}{2 G}r^2
\left\{\frac{8\pi G}{c^4}h_R \varepsilon -\frac{1}{2}\left(h-h_R R\right)-e^{-\lambda}\left[h_R^{\prime\prime}-\left(\frac{1}{2}\lambda^\prime-\frac{2}{r}\right)h_R^{\prime}\right]
\right\}_{(0)}.
\end{equation}
Here the presence of the subscript (0) by the curly brackets means that all functions in the brackets are taken from solutions of
Einstein's general relativistic equations.

Finally, the $i=r$ component
of the conservation law,
$T^k_{i; k}=0$, yields the  equation
\begin{equation}
\label{conserv_gen}
\frac{d p}{d r}=-\frac{1}{2}\left(\varepsilon+p\right)\frac{d \nu}{d r}.
\end{equation}

Altogether then, for obtaining background solutions in the MGT we will use Eqs.~\eqref{mod_Einstein-11_gen}, \eqref{mass_eq}, and \eqref{conserv_gen}
for three unknown functions $\nu, \lambda, p$. Note that in Eq.~\eqref{mod_Einstein-11_gen}, as well as in Eq.~\eqref{mass_eq},
the terms with $\alpha$ are calculated by using solutions of the Einstein equations.

Since the perturbative terms of Eqs.~\eqref{mod_Einstein-11_gen} and \eqref{mass_eq} contain the function of the scalar curvature $h(R)$,
it is convenient to rewrite $R$ through the trace of the energy-momentum tensor $T$,
\begin{equation}
\label{EMT_trace}
R_{(0)}=-\frac{8\pi G}{c^4} T_{(0)}\equiv -\frac{8\pi G}{c^4}(\varepsilon_{(0)}-3 p_{(0)}).
\end{equation}
This expression contains the pressure and the energy density determined from a solution of Eq.~\eqref{conserv_gen}
in GR using the corresponding EoS (see below).

\subsection{Magnetic field equations}
\label{mag_f_eqs}

Consistent with the statement of the problem given at the beginning of Sec.~\ref{statem_prob},
static spherically symmetric configurations described in Sec.~\ref{mod_eqs_gen}
will be used as a background for the magnetic field,
which is assumed to be created by a 4-current
$j_{\mu}=(0,0,0,j_{\phi})$~\cite{Konno:1999zv}. For such a current,
the electromagnetic 4-potential $A_{\mu}$
has only a $\phi$-component $A_{\mu}=(0,0,0,A_{\phi})$. In this case, taking into account the nonvanishing components of
the electromagnetic field tensor $F_{r \phi}=\partial A_{\phi}/\partial r$ and $F_{\Theta \phi}=\partial A_{\phi}/\partial \Theta$,
the Maxwell equations, written in the metric \eqref{metric_schw}, give
the following elliptic equation:
\begin{equation}
\label{maxw_A}
e^{-\lambda}
\frac{\partial^2 A_{\phi}}{\partial r^2}+\frac{1}{2} \left(\nu^\prime-\lambda^\prime\right)e^{-\lambda}\frac{\partial A_{\phi}}{\partial r}+
\frac{1}{r^2}\frac{\partial^2 A_{\phi}}{\partial \Theta^2}-
\frac{1}{r^2}\cot \Theta\frac{\partial A_{\phi}}{\partial \Theta}=-\frac{1}{c} j_{\phi}.
\end{equation}
Its solution is sought as an expansion of
 the potential $A_{\phi}$ and the current $j_{\phi}$ as follows~\cite{Regge:1957,Konno:1999zv}:
\begin{eqnarray}
\label{expan_A}
&& A_{\phi}=\sum_{l=1}^\infty a_l(r)\sin \Theta \frac{d P_l(\cos \Theta)}{d\Theta},\\
&& j_{\phi}=\sum_{l=1}^\infty j_l(r)\sin \Theta \frac{d P_l(\cos \Theta)}{d\Theta},
\end{eqnarray}
where $P_l$ is the Legendre polynomial of degree $l$. Substituting these expansions into Eq.~\eqref{maxw_A}, we have
\begin{equation}
\label{maxw_A_expan}
e^{-\lambda}
a_l^{\prime \prime}+\frac{1}{2}\left(\nu^\prime-\lambda^\prime\right)e^{-\lambda} a_l^\prime-\frac{l(l+1)}{r^2}a_l=-\frac{1}{c}j_l.
\end{equation}
Solution of this equation is sought for a given current~$j_l$, an equation for
which will be obtained below. Since here we consider only a dipole magnetic field
for which $l=1$, for convenience, we drop the subscript~1 on $a$ and~$j$ hereafter.

For the magnetic field under consideration,
the current $j$ is not arbitrary but has to satisfy an integrability condition~\cite{Bocquet:1995je,Konno:1999zv}.
To obtain this condition, we take into account the fact that the magnetic field induces only small deviations
in the shape of the background spherically symmetric configuration.
To describe  these deviations,
we follow the approach adopted in Ref.~\cite{Konno:1999zv} and expand
the metric in multipoles around the spherically symmetric
spacetime. Then one can regard
the deformations of the metric and of the fluid as second-order perturbations, whereas the electromagnetic field potential and the
current will be regarded as first-order perturbations. In this case the corresponding metric can be chosen in the form
\begin{eqnarray}
\label{pert_metr}
&& ds^2  =   e^{\nu (r)} \left\{ 1 + 2 \left[ h_{0}(r) + h_{2}(r)
          P_{2}( \cos \Theta ) \right] \right\} (dx^0)^2  
   -  e^{\lambda(r)}\left\{ 1 + \frac{2 e^{\lambda(r)}}{r}
          \left[ m_{0}(r) + m_{2}(r) P_{2}( \cos \Theta )
          \right] \right\} dr^2
          \nonumber \\
  & & -  r^2 \left[ 1 + 2 k_{2}(r) P_{2}( \cos \Theta )
        \right] \left( d \Theta^2
        +  \sin^2 \Theta d \phi^2 \right) \!,
\end{eqnarray}
where $h_0$, $h_2$, $m_0$, $m_2$, and $k_2$ are the second-order corrections of the metric
associated with the presence of the magnetic field, and $P_2$ denotes
the Legendre polynomial of order~2.

The total energy-momentum tensor for the system under consideration is
\begin{equation}
\label{EMT_total}
T_{i }^k=\left(\varepsilon +p\right)u^k u_i-\delta_i^k p
-F^k_\alpha F_i^\alpha+\frac{1}{4}\delta^k_i F_{\alpha \beta}F^{\alpha \beta}.
\end{equation}

Expand now the fluid energy density and pressure in the form
\begin{eqnarray}
\label{expans}
&&\varepsilon(r,\Theta)=\varepsilon_0+\frac{\varepsilon_0^\prime}{p_0^\prime}\left(\delta p_0+\delta p_2 P_2\right),\\
&&p(r,\Theta)=p_0+\delta p_0+\delta p_2 P_2,
\end{eqnarray}
where the background solutions $\varepsilon_0, p_0$ and the perturbations $\delta p_0, \delta p_2$
depend on $r$ only.
Substituting   these expansions into the conservation law, $T^k_{i; k}=0$,
and using the metric \eqref{pert_metr} and the background equation \eqref{conserv_gen},
we  obtain the following $i=r$ and $i=\Theta$ components (cf. Ref.~\cite{Konno:1999zv}):
\begin{eqnarray}
\label{p2expres1}
&&\delta p_2^\prime=-\left(\varepsilon_0+ p_0\right)h_2^\prime+
\frac{\varepsilon_0^\prime+p_0^\prime}{\varepsilon_0+p_0}\delta p_2
-\frac{2}{3}\frac{a^\prime}{r^2}\frac{j}{c},\\
\label{p2expres2}
&&\delta p_2=-\left(\varepsilon_0+p_0\right) h_2-\frac{2}{3}\frac{a}{r^2}\frac{j}{c}.
\end{eqnarray}

The integrability condition for Eqs.~\eqref{p2expres1} and \eqref{p2expres2} gives the equation for the current:
\begin{equation}
\label{ic_current}
j^\prime-\left(\frac{2}{r} +\frac{\varepsilon_0^\prime+p_0^\prime}{\varepsilon_0+p_0}\right)j=0,
\end{equation}
which can be integrated analytically to give
\begin{equation}
\label{ic_current_isot}
j=c_j r^2 (\varepsilon_0+p_0),
\end{equation}
where $c_j$ is an integration constant.
Thus the equation for the magnetic field \eqref{maxw_A_expan} contains the current determined by the expression \eqref{ic_current_isot}.

\section{Numerical results}
\label{num_calc}

In this section we numerically integrate the obtained equations for two special choices of the function $h(R)$.
In doing so, one needs some EoS for the neutron matter. This can be any of EoSs  used in the literature
in describing matter at high densities and pressures (see, for example, Ref.~\cite{Potekhin:2011xe}). Here we use the well-known
SLy EoS, which can be represented by the following analytical approximation~\cite{Haensel:2004nu}:
\begin{eqnarray}
\label{EOS_analyt}
\zeta=\frac{a_{1}+a_{2}\xi+a_{3}\xi^3}{1+a_{4}\xi}f(a_{5}(\xi-a_{6}))+(a_{7}+a_{8}\xi)f(a_{9}(a_{10}-\xi)) \nonumber\\
+(a_{11}+a_{12}\xi)f(a_{13}(a_{14}-\xi))+(a_{15}+a_{16}\xi)f(a_{17}(a_{18}-\xi))
\end{eqnarray}
with
$
\zeta=\log(p/\mbox{dyn}\, \mbox{cm}^{-2}), \, \xi=\log(\rho/\mbox{g}\,\mbox{cm}^{-3})\,,
$
where $\rho$ is the neutron matter density and $f(x)=[\exp(x)+1]^{-1}$.
The values of the coefficients $a_{i}$ can be found in Ref.~\cite{Haensel:2004nu}.

To model deviations from Einstein gravity, the literature in the field offers a number of functions $h(R)$
which are applied to modeling neutron stars and give a realistic description of mass-radius relations
for such type of compact objects. Examples of $h(R)$ can be found, for instance,  in Refs.~\cite{Arapoglu:2010rz,Alavirad:2013paa,Astashenok:2013vza}.
For our purpose, we employ two simple expressions~\cite{Astashenok:2013vza}:
\begin{itemize}
  \item
A power law, when the gravitational Lagrangian \eqref{Lagr_mod_pert} is chosen in the form
\begin{equation}
\label{h_R_part_pow}
f=R+\alpha h(R)\equiv R+\alpha R^2 \left(1+\gamma R\right).
\end{equation}
The values of the free parameters $\gamma$ and  $\alpha$ appearing here should be constrained from observations.
In the case of pure $R$-squared gravity (i.e., when $\gamma=0$) there are the following constraints on
 $\alpha$: (i)~in the weak-field limit, it is constrained by binary pulsar data as $\alpha \lesssim 5\times 10^{15} \text{cm}^2$~\cite{Naf:2010zy};
(ii) in the strong gravity regime, the constraint is $\alpha \lesssim 10^{10} \text{cm}^2$~\cite{Arapoglu:2010rz}. Since here, following Ref.~\cite{Astashenok:2013vza},
 we consider the case where
 $|\gamma R|\sim \mathcal{O}(1)$, the constraints on $\alpha$ can already differ from the case of
pure $R$-squared gravity. In this connection, the values of the parameters
 $\alpha$ and $\gamma$ will be chosen so as to satisfy the observations for the mass-radius relation given in Ref.~\cite{Ozel:2010fw}.
Notice that since here we employ the metric signature distinct from that of Ref.~\cite{Astashenok:2013vza}, below we take $\alpha, \gamma$
with opposite signs compared with~\cite{Astashenok:2013vza}.
  \item
An exponential law, when
\begin{equation}
\label{h_R_part_exp}
f=R+\beta h(R)\equiv R+\beta R \left[\exp{\left(-R/R_{\text{ch}}\right)}-1\right],
\end{equation}
where $R_{\text{ch}}$ is some characteristic value of the curvature.
Here instead of $\alpha$ we use  the dimensionless parameter $\beta$ to distinguish this case from that of given by Eq.~\eqref{h_R_part_pow}.
\end{itemize}

\subsection{Dimensionless equations and the procedure of solving}

For numerical calculations,
it is convenient to rewrite Eqs.~\eqref{mod_Einstein-11_gen}, \eqref{mass_eq}, \eqref{conserv_gen}, and \eqref{maxw_A_expan} in terms of dimensionless variables
\begin{equation}
\label{dimless_var_NS_mod}
x=\frac{r}{L}, \quad \Sigma=R L^2, \quad v(x)=\frac{M(r)}{4\pi 10^{\xi_c} L^3},  \quad \bar{j}=\frac{1}{10^{\xi_c/2}c^2} j,
\quad \bar{a}=\frac{8\pi G}{c^3}10^{\xi_c/2} a,
\end{equation}
where $L$ is some characteristic length (which is taken to be $L=10^6\,\text{cm}$ in the numerical calculations presented below)
and $\xi_c$ is the central density. Using these variables
and choosing $h(R)$ in the form of \eqref{h_R_part_pow},
we have from~\eqref{mod_Einstein-11_gen}, \eqref{mass_eq}, and~\eqref{conserv_gen}
\begin{eqnarray}
\label{conserv_part}
&&\xi^\prime=-\frac{1}{2 \ln{10}}\frac{1}{d\zeta/d\xi}\left(c^2 10^{\xi-\zeta}+1\right)\nu^\prime,
 \\
\label{mod_Einstein-11_part}
&&-(1-\mu)\left(\frac{1}{x^2}+\frac{\nu^\prime}{x}\right)+\frac{1}{x^2} \nonumber
\\
&&+\bar{\alpha}
\left[
-2 b \Sigma\left(1+\frac{3}{2} \bar{\gamma}\Sigma\right)10^\zeta+\frac{1}{2}\Sigma^2\left(1+2 \bar{\gamma}\Sigma\right)
-2 (1-\mu)\Sigma^\prime \left(1+3\bar{\gamma}\Sigma\right)\left(\frac{1}{2}\nu^\prime+\frac{2}{x}\right)
\right]_{(0)}=-b 10^\zeta,
 \\
\label{mass_eq_part}
&&v^\prime=x^2 10^{\xi-\xi_c}-2 \bar{\alpha} \Sigma_{(0)}
\Big[
x^2 10^{\xi-\xi_c}\left(1+\frac{3}{2} \bar{\gamma}\Sigma\right)+\frac{\delta}{4} x^2 \Sigma\left(1+2 \bar{\gamma}\Sigma\right)\nonumber
\\
&&+\left(\frac{1}{2}x^3 10^{\xi-\xi_c}-2\delta x+\frac{3}{2} v\right)\left(1+3\bar{\gamma}\Sigma\right)\frac{\Sigma^\prime}{\Sigma}-
x\left(\delta x-v\right)\frac{\Sigma^{\prime\prime}+3\bar{\gamma}\left(\Sigma^{\prime 2}+\Sigma \Sigma^{\prime\prime}\right)}{\Sigma}
\Big]_{(0)},
\end{eqnarray}
where the prime denotes now differentiation with respect to $x$, $\delta=c^2/\left(8\pi G L^2 10^{\xi_c}\right)$,  $\mu=v/(\delta \,  x)$, $\bar{\alpha}=\alpha/L^2, \bar{\gamma}=\gamma/L^2$, $b=8\pi G L^2/c^4$.
In the same way, one can derive dimensionless equations for $h(R)$ from Eq.~\eqref{h_R_part_exp}
(we do not show them here to avoid overburdening the text).

These equations are to be solved subject to the boundary conditions given in the neighborhood of the center by the following expansions:
\begin{equation}
\label{bound_mod_Ein}
\xi\approx\xi_c+\frac{1}{2}\xi_2 x^2, \quad \nu\approx\nu_c+\frac{1}{2}\nu_2 x^2, \quad v\approx \frac{1}{3} v_3 x^3,
\end{equation}
where the expansion coefficients are determined from Eqs.~\eqref{conserv_part}-\eqref{mass_eq_part}.

The step-by-step procedure for finding solutions is as follows.
The first step is to find unperturbed solutions to Eqs.~\eqref{conserv_part}-\eqref{mass_eq_part}, i.e., to consider the case where
$\bar{\alpha}$ is set equal to zero. These solutions correspond to solutions of GR, and we denote them by the subscript
 (0), i.e.,
$\xi_{(0)}, \nu_{(0)}$, and $v_{(0)}$. Using these solutions, the dimensionless scalar curvature from
 \eqref{EMT_trace} is evaluated as
$$
\Sigma_{(0)}=-b \left(c^2 10^{\xi_{(0)}}-3\cdot 10^{\zeta_{(0)}}\right).
$$

In the second step, the obtained solutions are used in the square brackets
 $[\ldots]_{(0)}$ in Eqs.~\eqref{mod_Einstein-11_part} and \eqref{mass_eq_part}, that permits us to get solutions of the perturbed equations
 in the MGT. Let us denote these solutions by the subscript 0,  i.e., $\xi_{0}, \nu_{0}$, and $v_{0}$,
with a view to applying them as background solutions  in calculating the magnetic field by using the equations from Sec.~\ref{mag_f_eqs}.

Note that in making the computations
the integration is performed  from the center to the point $x_b$,
where the neutron matter density decreases to the value
 $\rho_b \approx 10^6 \text{g cm}^{-3}$.
 We take this point to be a boundary of the star~\cite{Orellana:2013gn}.
This density corresponds to the outer boundary of a neutron star crust
up to which the SLy EoS \eqref{EOS_analyt} remains valid~\cite{Haensel:2004nu}.


Finally, in the third step, the obtained background solutions are used for calculating the distribution of the magnetic field along the
radius of the configuration. To do this, rewrite Eqs.~\eqref{maxw_A_expan} and \eqref{ic_current_isot}
 in terms of the above dimensionless variables:
\begin{equation}
\label{maxw_A_expan_part}
(1-\mu_0)\left[
\bar{a}^{\prime \prime}+\frac{1}{2}\left(\nu_0^\prime -\frac{\mu_0^\prime}{1-\mu_0}\right)\bar{a}^\prime
\right]-\frac{2}{x^2}\bar{a}=-\bar{c}_j x^2\left(c^2 10^{\xi_0}+10^{\zeta_0}\right),
\end{equation}
where $\bar{c}_j$ is the rescaled arbitrary constant from \eqref{ic_current_isot} and $\mu_0(x)=v_0(x)/(\delta \,  x)$.
The boundary condition for this equation is given
by an expansion in the neighborhood of the center,
\begin{equation}
\label{bound_cond_magn}
\bar{a} \approx \frac{1}{2}\, a_c x^2,
\end{equation}
where  $a_c$ is a free parameter. This parameter and the arbitrary constant $\bar{c}_j$
are chosen in such a way as to (i)~get a required value of the surface magnetic field, and
(ii)~obtain asymptotically decaying solutions for $x \rightarrow \infty$.

There is known external solution
for the electromagnetic field potential, according to which beyond the fluid
$\bar{a}$ decays as~\cite{Konno:1999zv}
$$
\bar{a}\sim -x^2\left[\ln{(1-\mu_0)}+\mu_0+\frac{1}{2}\mu_0^2\right].
$$
In this solution,  $\mu_0$ corresponds to the external vacuum solution for the background configuration with the mass concentrated
inside the radius $x_b$.

Finally,
for the strength of the magnetic field, one can derive
the following tetrad components  (i.e., the components
measured by a locally inertial observer):
\begin{equation}
\label{streng_magn_dmls_NS_mod}
B_{\hat{r}}=-F_{\hat{\Theta} \hat{\phi}}= \frac{2 c^3}{8\pi G L^2 10^{\xi_c/2}}\frac{\bar{a}}{x^2}\cos{\Theta}, \quad
B_{\hat{\Theta}}=F_{\hat{r} \hat{\phi}}=- \frac{c^3}{8\pi G L^2 10^{\xi_c/2}}\frac{\sqrt{1-\mu}}{x}\,\bar{a}^\prime\sin{\Theta}.
\end{equation}

\subsection{Structure of the magnetic field}
\label{mag_f_str}

The structure of the magnetic field is calculated for a set of configurations in the MGT with the functions
\eqref{h_R_part_pow} and \eqref{h_R_part_exp}. As pointed out in Ref.~\cite{Astashenok:2013vza},
for models with these functions, there are two branches of stable solutions, i.e., solutions for which
$dM/d\rho_c>0$, where $\rho_c$ is the central density of neutron matter. One branch lies approximately  in the same
range of  $\rho_c$ where the solutions of GR are located.
Masses and sizes of such configurations are very close to those of systems in GR, and it can be shown that the structure
of the magnetic field is in practice indistinguishable from that of neutron stars in GR.
In order to not encumber the paper, here we do not consider configurations lying on this stable branch.

However, in contrast to GR, there exists another branch of stable solutions
which lies at greater central densities.
Our purpose here is to study magnetic fields of configurations in the MGT which lie on this stable branch and to compare them
to those of neutron stars in GR.
The comparison is performed as follows:
\begin{itemize}
  \item
We consider pairs of configurations, each of which contains systems constructed in GR and in the MGT. The configurations lie on two different stable branches,
have different radii but the same masses.
  \item
  The free parameters of the magnetic field
$a_c, \bar{c}_j$ are  chosen  in such a way as to provide the required surface magnetic field $B_s$ (say, at the pole) for all configurations,
and also to obtain asymptotically vanishing solutions for the electromagnetic potential.
\end{itemize}

\begin{figure}[t]
\centering
  \includegraphics[height=10cm]{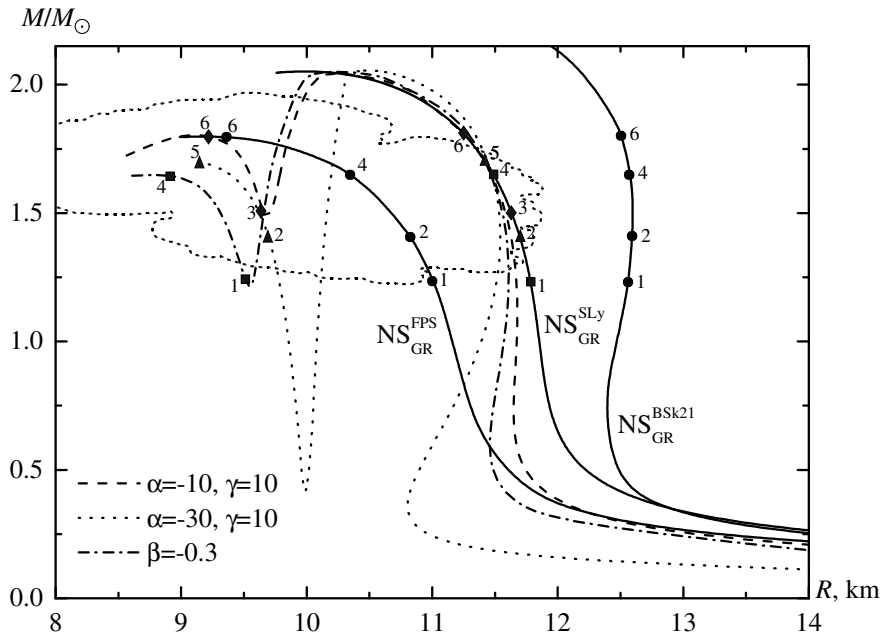}
\vspace{-1cm}
\caption{The mass-radius relations for neutron stars in GR (denoted as $\text{NS}_{\text{GR}}^{\text{SLy}}$,
$\text{NS}_{\text{GR}}^{\text{FPS}}$, and $\text{NS}_{\text{GR}}^{\text{BSk21}}$ for the SLy, FPS, and BSk21 EoSs, respectively)
and in the $f(R)$ models \eqref{h_R_part_pow} and \eqref{h_R_part_exp} obtained for the SLy
EoS \eqref{EOS_analyt}. Here and in Fig.~\ref{fig_magn_f_power}
the values of the parameter $\alpha$ are given in units of $10^9 \text{cm}^2$ and of $\gamma$ -- in units of $\left(r_{g \odot}/2\right)^2$~\cite{Astashenok:2013vza},
where $r_{g \odot}$ is the gravitational radius of the Sun.
The dotted contour depicts the region of the observational constraints~\cite{Ozel:2010fw}.
The numbers near the bold symbols
correspond to the configurations with the same masses equal to: for  1 -- $1.24 M_{\odot}$, for  2 -- $1.41 M_{\odot}$,
for 3 -- $1.50 M_{\odot}$, for 4 -- $1.65 M_{\odot}$, for 5 -- $1.70 M_{\odot}$, for 6 -- $1.80 M_{\odot}$.
The comparison is carried out for the configurations 3 and 6 [the model \eqref{h_R_part_pow} with $\alpha=-10, \gamma=10$],
2 and 5 [the model \eqref{h_R_part_pow} with $\alpha=-30, \gamma=10$], 1 and 4 [the model \eqref{h_R_part_exp} with $\beta=-0.3$]
(see in the text).
}
\label{fig_M_R_relat}
\end{figure}

For this purpose,  we constructed mass-radius relations for neutron stars in GR and in the MGT shown in Fig.~\ref{fig_M_R_relat}.
Here the dotted contour depicts the region of observational constraints obtained for
three neutron stars~\cite{Ozel:2010fw}.
Taking into account these constraints, we chose the pairs of configurations (shown by the bold symbols in Fig.~\ref{fig_M_R_relat}) so that they
were located either in the mentioned region or very close to it.

Notice that a necessary condition for the validity of the perturbative approach used here
is that the maximum value of the quantity $\Delta=\left|\alpha h(R)/R\right|$ would be always
$\Delta_{\text{max}}\lesssim 0.1$ along the radius.
For the configurations under consideration, this condition is slightly violated only in the case
of the model \eqref{h_R_part_pow} with $\alpha=-30, \gamma=10$, when for the configuration with
$M=1.70 M_{\odot}$ we have $\Delta_{\text{max}} \approx 0.19$ at the center.
In all the remaining cases considered here the condition for the perturbative approach to be valid is satisfied with sufficient accuracy.

\begin{figure}[t]
\begin{minipage}[t]{.49\linewidth}
  \begin{center}
  \includegraphics[width=7.cm]{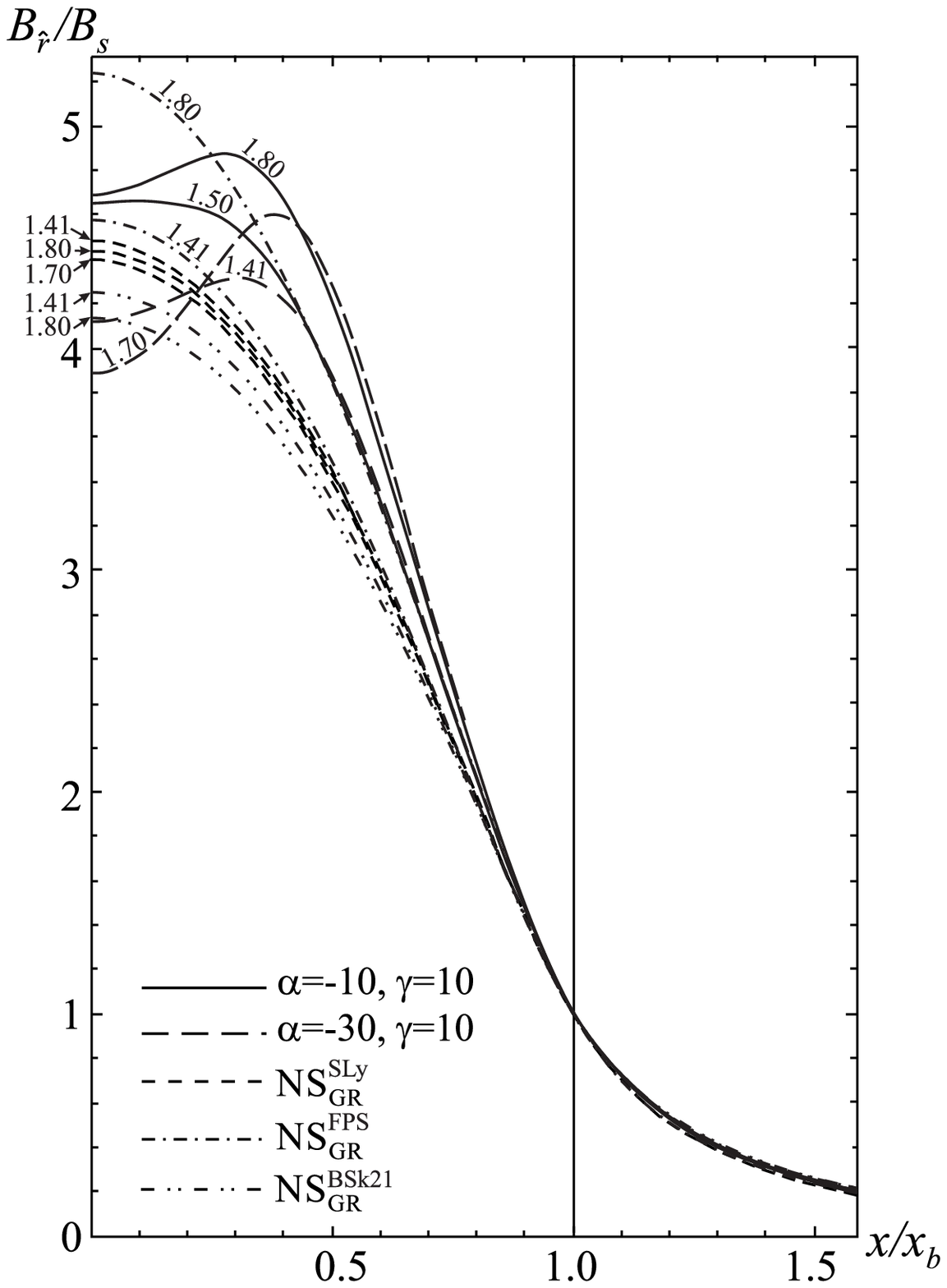}
  \end{center}
\end{minipage}\hfill
\begin{minipage}[t]{.49\linewidth}
  \begin{center}
  \includegraphics[width=7.1cm]{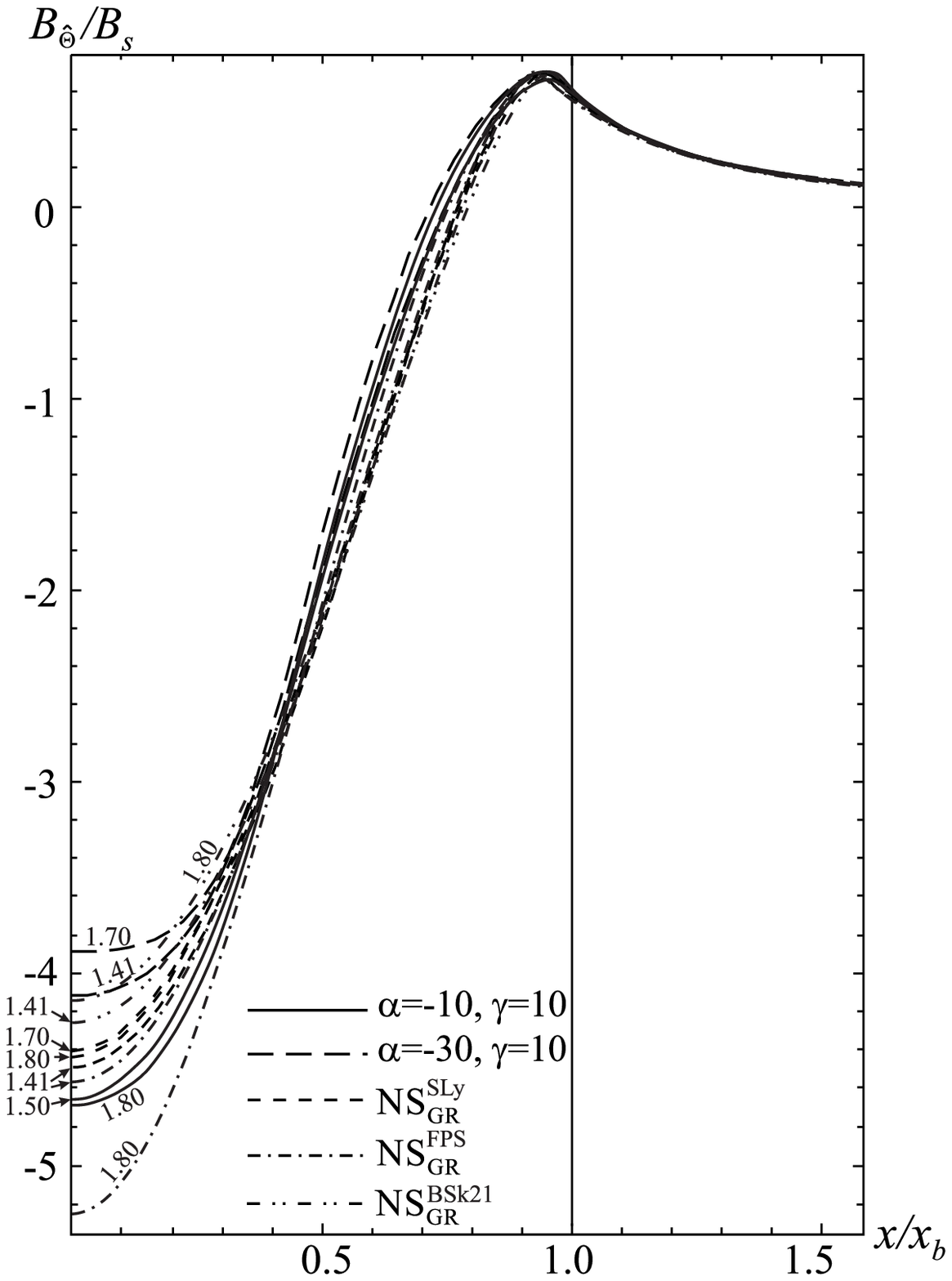}
  \end{center}
\end{minipage}\hfill
\caption{
The tetrad components $B_{\hat{r}}$ and $B_{\hat{\Theta}}$
of the magnetic field (in units of the surface strength
of the magnetic field $B_s$ at the pole) in GR and in the $f(R)$ model \eqref{h_R_part_pow} evaluated
on the symmetry axis ($\Theta=0$) and
in the equatorial plane ($\Theta=\pi/2$), respectively,
are shown as functions of the relative radius $x/x_b$.
The curves are plotted for the configurations shown in Fig.~\ref{fig_M_R_relat}
by the bold symbols.
The thin vertical lines correspond to the boundary of the fluid.
The numbers near the curves denote masses of the configurations (in solar mass units).
}
\label{fig_magn_f_power}
\end{figure}

\begin{figure}[h!]
\begin{minipage}[t]{.49\linewidth}
  \begin{center}
  \includegraphics[width=6.9cm]{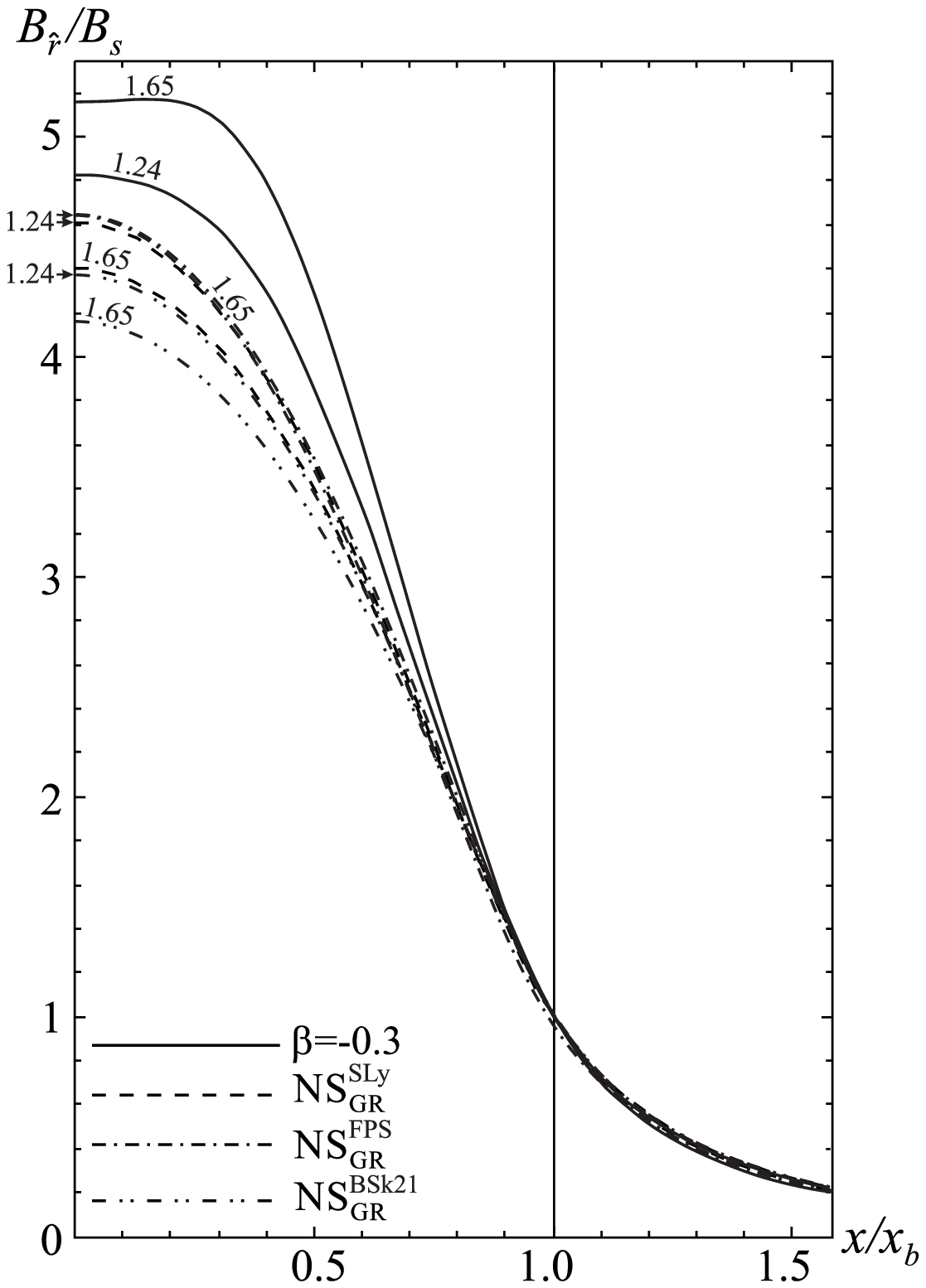}
  \end{center}
\end{minipage}\hfill
\begin{minipage}[t]{.49\linewidth}
  \begin{center}
  \includegraphics[width=7cm]{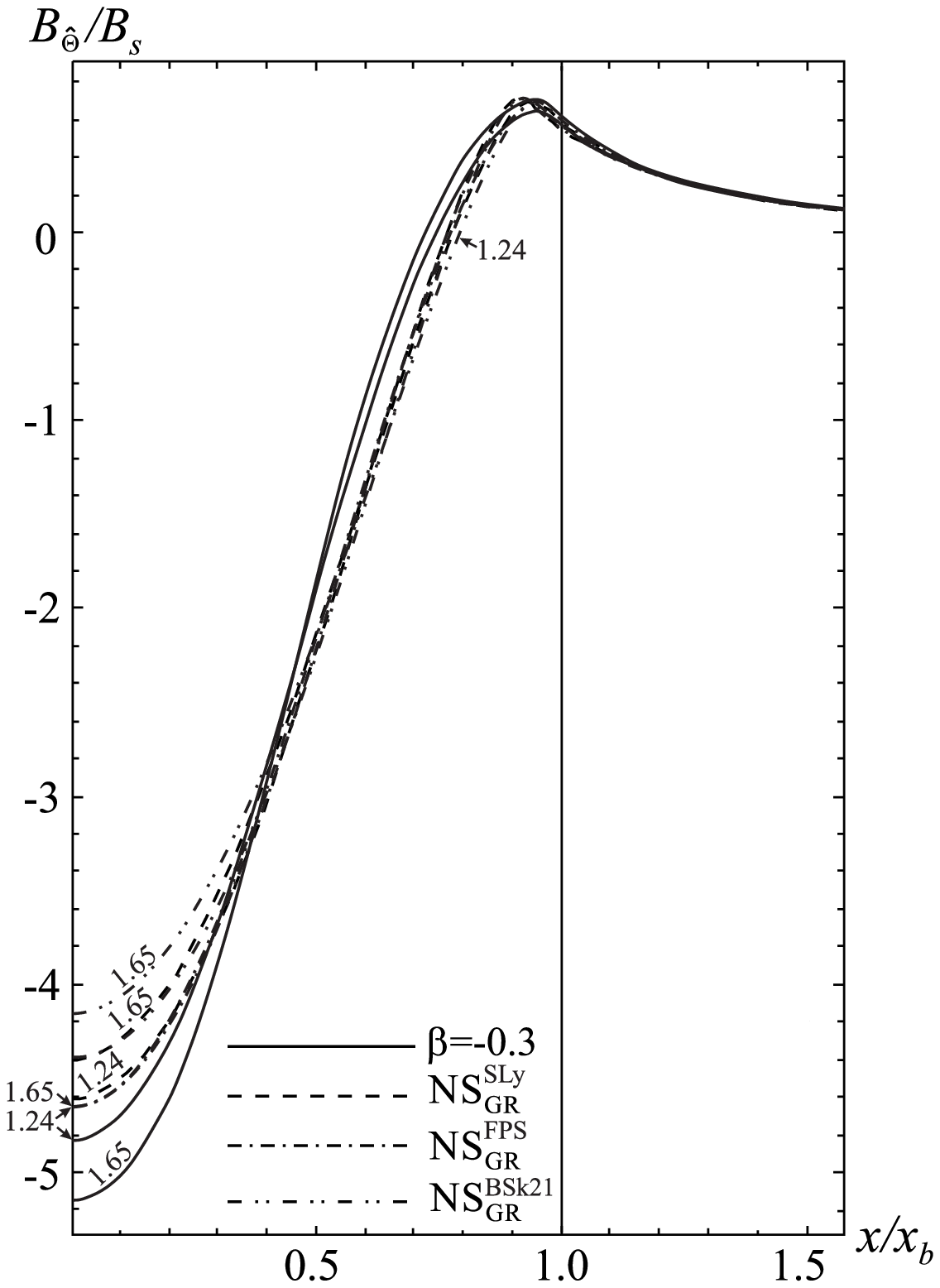}
  \end{center}
\end{minipage}\hfill
\caption{The same as in Fig.~\ref{fig_magn_f_power} for the model \eqref{h_R_part_exp}. The constant $R_{\text{ch}}=-0.5\, \left(r_{g \odot}/2\right)^{-2}$~\cite{Astashenok:2013vza}.
}
\label{fig_magn_f_exp}
\end{figure}

Using the expressions \eqref{streng_magn_dmls_NS_mod},
we calculated the components of the magnetic field for the configurations of Fig.~\ref{fig_M_R_relat}, which are
shown in Figs.~\ref{fig_magn_f_power} and \ref{fig_magn_f_exp}.
For purposes of comparison, the graphs are plotted using relative units where the current radius
$x$ is normalized to the radius of the fluid $x_b$, and the magnetic field is measured
in units of the surface strength $B_s$ at the pole.

It is known from the literature that the structure
of the interior magnetic field of neutron stars may depend substantially on the choice of EoS
(see, e.g., Refs.~\cite{Bocquet:1995je,Kiuchi:2007pa}). Therefore,
for comparing changes induced by the effects of modified gravity
over the changes arising when one modifies an EoS only,
we have also considered the structure of the magnetic field
of neutron stars in GR when the fluid is modeled
by two other  EoSs (apart from the SLy EoS): the~softer
FPS EoS~\cite{Haensel:2004nu}  and the stiffer BSk21 EoS~\cite{Potekhin:2013qqa}. The corresponding mass-radius curves are shown in Fig.~\ref{fig_M_R_relat},
and the distributions of the magnetic field for these EoSs -- in Figs.~\ref{fig_magn_f_power} and \ref{fig_magn_f_exp} for the configurations
  1,~2,~4,~6 of Fig.~\ref{fig_M_R_relat}.

It is seen that the magnitude and the distribution of the interior magnetic field
change substantially depending on the specific choice of $h(R)$ and on the values of the free parameters here used.
In particular, for the case of $h(R)$ from \eqref{h_R_part_pow} shown in Fig.~\ref{fig_magn_f_power}
 the component $B_{\hat{r}}$  may either be always
greater  than that of the GR case or have a mixed behavior when it is less in the central region
and then becomes greater than that of the GR case. In turn, the behavior of the component $B_{\hat{\Theta}}$ also depends appreciably on
the chosen values of the parameters of the systems under consideration.

 As for the influence of changes in an EoS, one can see from Figs.~\ref{fig_magn_f_power} and \ref{fig_magn_f_exp} that
 there are the following
 modifications in the behaviour of the field along the radius of the system:
 (i)  for the stiffer EoS (BSk21), the component $B_{\hat{r}}$ is smaller (modulus),
 and for the softer EoS (FPS) -- is greater, compared with the case of the SLy EoS;
 (ii) the magnitude of the component $B_{\hat{\Theta}}$ depends on the radius where the comparison is performed,
 and it may either be greater or less than that of the case of the SLy EoS.
  But the modifications of the field,
 whether they are associated with changes in an EoS or
 are appeared due to the modification of gravity, are comparable to each other in magnitude. So one may conclude that the effects of modification of gravity are just as important
 in their impact on the structure of the magnetic field
 as are the effects associated with changes in an EoS in modeling
 the distribution of the magnetic field within the framework of GR.

\begin{figure}[t]
\centering
  \includegraphics[height=16cm]{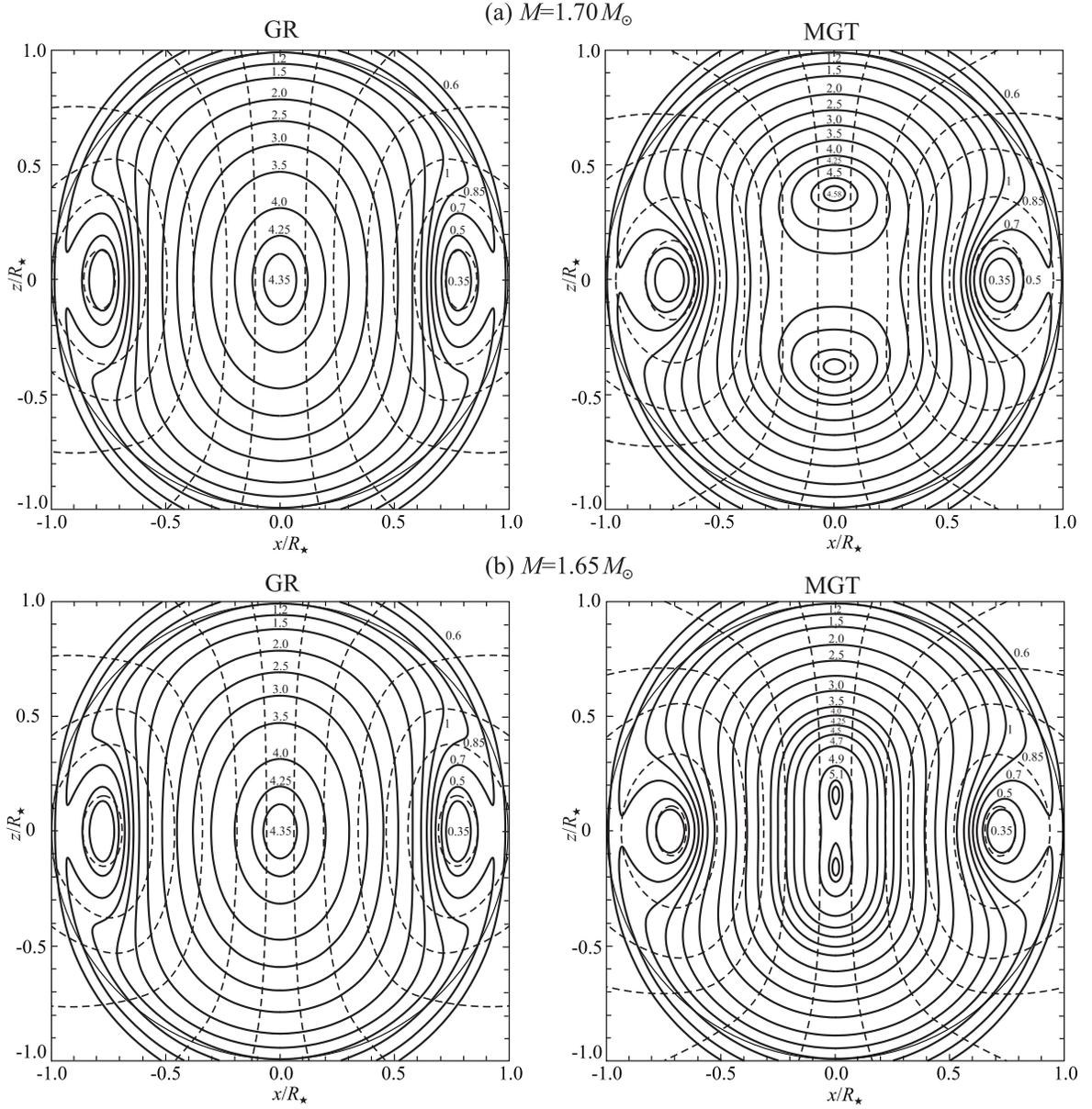}
\caption{Equipotential lines of the total
magnetic field strength $B$ (the solid lines) and magnetic field lines (the dashed lines)
for the configurations 5 and 4 of Fig.~\ref{fig_M_R_relat} with the SLy EoS.
The figure shows plots for the neutron stars
in the $f(R)$ model \eqref{h_R_part_pow} with $\alpha=-30, \gamma=10$ [Fig.~(a)]
and in the model \eqref{h_R_part_exp} with $\beta=-0.3$ [Fig.~(b)].
The plots are made in a meridional plane $\phi=\text{const.}$ spanned by the coordinates
$x=r \sin{\Theta}$ and $z=r \cos{\Theta}$. The numbers near the solid curves correspond to the magnitude of
$B$ in units of $B_s$ at the pole.
The thin-line circles denote the boundary of the neutron fluid,
possessing the radius $R_\star$.
}
\label{fig_force_lines}
\end{figure}

 It is also interesting to note that, independently of EoSs used here, the qualitative behaviour of the magnetic field of the neutron stars in GR
 does not change.
 A different situation takes place in the MGT, where there is
 a new type of behavior
of the component $B_{\hat{r}}$, which may already have a maximum not at the center (as it is for the neutron stars in GR)
but somewhere in the intermediate region between the center and the edge of a star.
As a result, the force lines of the magnetic field of the neutron stars in GR and in the MGT have a qualitatively different structure
in the internal regions of the configurations.
The situation is illustrated by  Fig.~\ref{fig_force_lines}, which shows the typical distributions of
the force lines: if the maximum value of the field strength of the systems in GR
is always located at the center, the maximum of the field for the neutron stars in the MGT is shifted away from the center.

Note that Eqs.~\eqref{maxw_A_expan} and \eqref{ic_current} are invariant under
the transformation $a, j \to \kappa a, \kappa j$,  $\kappa$ being any constant.
Correspondingly, the components
of the magnetic field given by Eq.~\eqref{streng_magn_dmls_NS_mod} transform as
 $B_{\hat{r}}, B_{\hat{\Theta}}\to \kappa B_{\hat{r}}, \kappa B_{\hat{\Theta}}$. Then, if one simultaneously replaces
 $B_s$ by $\kappa B_s$, the graphs shown in Figs.~\ref{fig_magn_f_power}-\ref{fig_force_lines}  remain unchanged for any value of the field, and the
dimensional values (in gauss) of the strength of the magnetic field are obtained on multiplying these graphs
by the required surface value $B_s$.
Of course, this holds true only within the perturbative approximation used here when one can neglect the influence of the magnetic field on the background configurations.
In particular, the obtained graphs can be applied both to the ``classical pulsars''
(for which $B_s\sim 10^{12}$~G) and to
magnetars (for which $B_s\sim 10^{15}$~G).

\section{Ellipticity}
\label{ellipticity}

\begin{figure}[t]
\centering
  \includegraphics[height=9cm]{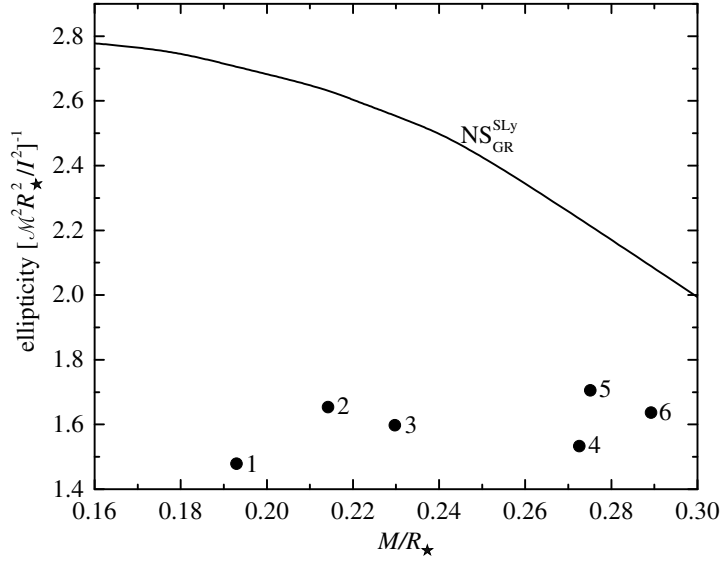}
\vspace{-1cm}
\caption{Ellipticity for the configurations with the SLy EoS. The bold dots denote  systems in the MGT, whose masses are given in the caption of Fig.~\ref{fig_M_R_relat}.
The ellipticity for the configurations 1, 4 is calculated in the model \eqref{h_R_part_exp} with $\beta=-0.3$; for the systems 3, 6~-- in the model \eqref{h_R_part_pow} with $\alpha=-10, \gamma=10$;
 for the systems 2, 5~-- in the model \eqref{h_R_part_pow} with $\alpha=-30, \gamma=10$.
}
\label{fig_ellipt}
\end{figure}

The presence of the axisymmetric magnetic field results in a deviation of the shape of the configurations from spherical symmetry.
As pointed out at the beginning of Sec.~\ref{statem_prob}, such deviations are small for magnetic field strengths considered in the present paper.
Nevertheless, it is of some interest to compare these deviations for neutron stars in GR and in the MGT.

To do so, following Ref.~\cite{Konno:1999zv}, let us introduce the ellipticity of a star
\begin{equation}
\label{ellipt}
\text{ellipticity}\equiv\frac{\text{(equatorial radius)--(polar radius)}}{\text{(mean radius)}}
=\frac{2 c_j}{r \nu_0^\prime}a+\frac{3 h_2}{r \nu_0^\prime}-\frac{3}{2}k_2.
\end{equation}
The physical meaning of the components on the right-hand side of this expression is as follows.
The first term corresponds to the Lorentz force, the second one describes the perturbation of the gravitational potential
arising due to the presence of the magnetic field, and the third one -- the perturbation of the circumferential radius.

To calculate the functions $h_2$ and $k_2$ appearing in Eq.~\eqref{ellipt}, let us employ the following equations
 \cite{Konno:1999zv} (see also in Ref.~\cite{Folomeev:2015aua}):
\begin{eqnarray}
\label{pert_Einst_G21_l2}
&& h_2^\prime+k_2^\prime=\left(\frac{1}{r}-\frac{1}{2}\nu_0^\prime\right)h_2+\left(\frac{1}{r}+\frac{1}{2}\nu_0^\prime\right)\frac{m_2}{r(1-\mu_0)}+
\frac{16\pi G}{3 c^4}\frac{a a^\prime}{r^2}, \\
\label{pert_Einst_G00_l2}
 && k_2^{\prime\prime}-\left[\frac{\mu_0^\prime}{2(1-\mu_0)}-\frac{3}{r}\right]k_2^\prime
-\frac{1}{r^2(1-\mu_0)}\left[2 k_2+m_2^\prime+\frac{3 m_2}{r(1-\mu_0)}\right] \nonumber\\
&&=\frac{4\pi G}{c^4}\frac{1}{1-\mu_0}\left(-\frac{\varepsilon_0^\prime}{p_0^\prime}\delta p_2
+\frac{1-\mu_0}{3}\frac{a^{\prime 2}}{r^2}-\frac{4}{3}\frac{a^2}{r^4}\right).
\end{eqnarray}
Here the function $m_2$ is defined by the expression
$$
 m_2=r(1-\mu_0)\left[-h_2+\frac{8\pi G}{3 c^4}(1-\mu_0)a^{\prime 2}\right],
$$
and $\delta p_2$ is taken from Eq.~\eqref{p2expres2}. To obtain regular solutions,
we choose the boundary conditions for these equations in the form
\begin{equation}
\label{bound_cond_pert}
h_2 \approx \frac{1}{2}h_{2c}r^2,
\quad k_2 \approx \frac{1}{2}k_{2c}r^2,
\end{equation}
where the expansion parameters appearing here are constrained as follows~\cite{Folomeev:2015aua}:
$$
h_{2c}+k_{2c}=\frac{2}{3}a_c^2.
$$

Using the dimensionless variables \eqref{dimless_var_NS_mod}, we numerically solved Eqs.~\eqref{conserv_part}-\eqref{mass_eq_part},
\eqref{maxw_A_expan_part}, \eqref{pert_Einst_G21_l2}, and \eqref{pert_Einst_G00_l2}. The results for the ellipticity~\eqref{ellipt}
are shown in Fig.~\ref{fig_ellipt}. To compare our results with those of Ref.~\cite{Konno:1999zv}, we normalized the ellipticity by
${\cal M}^2 R_\star^2/I^2$ and employed  geometric units $c=G=1$.
Here $I$ is the moment of inertia of a star of radius $R_\star$, ${\cal M}=B_s R_\star^3$ is the typical magnetic dipole moment.

One can see from Fig.~\ref{fig_ellipt} that,
in contrast to neutron stars with a polytropic EoS of Ref.~\cite{Konno:1999zv},
the ellipticity of the neutron stars with the SLy EoS in GR considered here
becomes smaller as the relativistic factor $M/R_\star$ increases.
In turn, it is seen that for a fixed $M/R_\star$ the ellipticity of the neutron stars
in the MGT is always smaller than that in~GR.

\section{Conclusion}
\label{conclusion}

We have studied  equilibrium, gravitating configurations within the framework of $f(R)$ gravity
consisting of a strongly magnetized neutron fluid.
In doing so, two simplifying approaches have been used: (i) The energy density of the magnetic field
(modeled here in the form of a dipole field) was assumed to be much smaller than the energy density of the neutron matter.
The axisymmetric deformations of a star associated with the dipole field
are small that allows one to consider them as
second-order corrections
to the background configurations and to employ the perturbative approach of Ref.~\cite{Konno:1999zv} in describing such systems.
(ii) As the background configurations, neutron stars modeled
in perturbative $f(R)$ gravity  have been used.
As shown in Ref.~\cite{Astashenok:2013vza}, such configurations, in contrast to  GR, may possess not just one but two branches of stable solutions,
and for the SLy EoS the new branch can describe the observational data even better than this is done in GR.

Working within the framework of these perturbative approaches,
our goal was
to clarify the question of how the presence of higher-order corrections to Einstein's GR may influence the structure of the magnetic field.
For this purpose, we compared configurations with equal masses from GR and from the MGT  lying on different branches of stability.
The main results of the studies can be summarized as follows:
\begin{enumerate}
\itemsep=-0.2pt
\item[(i)] Neutron stars have been modeled using the SLy EoS \eqref{EOS_analyt} and two types of the function $f(R)$:
  in the form of a power law~\eqref{h_R_part_pow} and in the exponential form~\eqref{h_R_part_exp}.
 For these models, the mass-radius relations have been constructed
 (see Fig.~\ref{fig_M_R_relat}), using which we have chosen the pairs of configurations (each of which contains systems from GR and from the MGT)
 satisfying the current observational constraints.

\item[(ii)] For these pairs, it was shown that the distribution of the interior magnetic field depends
substantially on the specific choice of $h(R)$ and on the values of the free parameters here used
(see Figs.~\ref{fig_magn_f_power}-\ref{fig_force_lines}). At the same time,
from the point of view of a distant observer,
configurations which enter in each of the pairs
possess the same masses and surface strengths of the magnetic field, and also have similar distributions of the external
magnetic field.

\item[(iii)] As was shown earlier in modeling neutron stars in  $f(R)$ gravity (see, e.g., Refs.~\cite{Orellana:2013gn,Alavirad:2013paa}),
the deviations from GR manifest themselves mainly near the surface of a star.
This can affect, in particular, the redshift  of spectral lines
of the star's atmosphere~\cite{Alavirad:2013paa}.
On the other hand, we have shown here that the magnetic field, on the contrary,
differs appreciably from GR mainly in the internal regions of a star, whereas its behavior
near the surface is practically the same as that of stars from GR.

\item[(iv)] Using two other EoSs (the softer FPS EoS and the stiffer BSk21 EoS), we have considered the structure of the magnetic field
of neutron stars in GR for comparing changes induced by the effects of modified gravity
over the changes arising when one modifies an EoS only.
It was shown that the effects of modification of gravity are just as important
 in their impact on the structure of the magnetic field
 as are the effects associated with changes in an EoS in modeling
 the distribution of the magnetic field within the framework of GR.

\item[(v)] The ellipticity of the configurations arising  due to the presence of the magnetic field has been calculated.
It was shown that for a fixed value of the relativistic factor $M/R_\star$
the ellipticity of neutron stars in GR is always larger than that of systems in the MGT (see Fig.~\ref{fig_ellipt}).
\end{enumerate}

Thus it is seen that the structure
of the interior magnetic field may depend substantially
not only on the physical properties of neutron matter
(given by different choices of EoSs,
as demonstrated, for instance,
in Refs.~\cite{Bocquet:1995je,Kiuchi:2007pa} and here, see Figs.~\ref{fig_magn_f_power} and \ref{fig_magn_f_exp}) and on the form of internal electric currents
but also on  which type of gravitational theory is used in modeling neutron stars.
Note also that these differences already manifest themselves within the framework of the perturbative approaches
used here in describing the modified gravity and magnetic fields.
Beyond the framework of these approaches, one may expect the appearance of
new changes in the structure of magnetic fields associated with the influence of modifications of Einstein's GR.
This is a possible subject for further investigation.

\vspace{-.2cm}
\section*{Acknowledgments}

V.F. gratefully acknowledges support provided by the Volkswagen Foundation
and by Grant No.~$\Phi.$0755 in fundamental research in natural sciences
of the Ministry of Education and Science of Kazakhstan.
We are  very grateful to  V.~Dzhunushaliev for fruitful discussions  and comments.
We would also like to thank anonymous referees for their valuable suggestions
 that have helped to improve the paper.


\begin{thebibliography}{999}
\bibitem{Potekhin:2011xe}
  A.~Y.~Potekhin,
  Phys.\ Usp.\  {\bf 53}, 1235 (2010)
  [Usp.\ Fiz.\ Nauk {\bf 180}, 1279 (2010)]
  [arXiv:1102.5735 [astro-ph.SR]].

\bibitem{MTG_early}
T. V. Ruzmaikina and A. A. Ruzmaikin, Zh. Eksp. Teor. Fiz. {\bf 57}, 680 (1969) [Sov. Phys. JETP {\bf 30}, 372 (1970)];\\
B. N. Breizman, V. Ts. Gurovich, and V. P. Sokolov, Zh. Eksp. Teor. Fiz. {\bf 59}, 288 (1970) [Sov. Phys. JETP {\bf 32}, 155
(1971)].

\bibitem{Nojiri:2010wj}
  S.~Nojiri and S.~D.~Odintsov,
  Phys.\ Rept.\  {\bf 505}, 59 (2011)
  [arXiv:1011.0544 [gr-qc]].

\bibitem{thick_b}
S.~Nojiri and S.~D.~Odintsov,
  JHEP {\bf 0007}, 049 (2000)
  [arXiv:hep-th/0006232];
I.~P.~Neupane,
  JHEP {\bf 0009}, 040 (2000)
  [arXiv:hep-th/0008190];
K.~A.~Meissner and M.~Olechowski,
  Phys.\ Rev.\  D {\bf 65}, 064017 (2002)
  [arXiv:hep-th/0106203];
  S.~Nojiri, S.~D.~Odintsov, and S.~Ogushi,
  Phys.\ Rev.\  D {\bf 65}, 023521 (2002)
  [arXiv:hep-th/0108172];
  M.~Parry, S.~Pichler, and D.~Deeg,
  JCAP {\bf 0504}, 014 (2005)
  [arXiv:hep-ph/0502048];
  V.~Dzhunushaliev, V.~Folomeev, B.~Kleihaus, and J.~Kunz,
  JHEP {\bf 1004}, 130 (2010)
  [arXiv:0912.2812 [gr-qc]].

\bibitem{Capozziello:2004us}
  S.~Capozziello, V.~F.~Cardone, S.~Carloni, and A.~Troisi,
  Phys.\ Lett.\ A {\bf 326}, 292 (2004)
  [gr-qc/0404114].

\bibitem{rel_star_f_R}
  A.~Upadhye and W.~Hu,
  Phys.\ Rev.\ D {\bf 80}, 064002 (2009)
  [arXiv:0905.4055 [astro-ph.CO]];
  E.~Babichev and D.~Langlois,
  Phys.\ Rev.\ D {\bf 81}, 124051 (2010)
  [arXiv:0911.1297 [gr-qc]].

\bibitem{wh_f_R}
 K.~A.~Bronnikov, M.~V.~Skvortsova, and A.~A.~Starobinsky,
  Grav.\ Cosmol.\  {\bf 16}, 216 (2010)
  [arXiv:1005.3262 [gr-qc]];
  A.~DeBenedictis and D.~Horvat,
  Gen.\ Rel.\ Grav.\  {\bf 44}, 2711 (2012)
  [arXiv:1111.3704 [gr-qc]].

\bibitem{Cooney:2009rr}
  A.~Cooney, S.~DeDeo, and D.~Psaltis,
  Phys.\ Rev.\ D {\bf 82}, 064033 (2010)
  [arXiv:0910.5480 [astro-ph.HE]].

\bibitem{Arapoglu:2010rz}
  A.~S.~Arapoglu, C.~Deliduman, and K.~Y.~Eksi,
  JCAP {\bf 1107}, 020 (2011)
  [arXiv:1003.3179 [gr-qc]].

\bibitem{Orellana:2013gn}
  M.~Orellana, F.~Garcia, F.~A.~Teppa Pannia, and G.~E.~Romero,
  Gen.\ Rel.\ Grav.\  {\bf 45}, 771 (2013)
  [arXiv:1301.5189 [astro-ph.CO]].

\bibitem{Alavirad:2013paa}
  H.~Alavirad and J.~M.~Weller,
  Phys.\ Rev.\ D {\bf 88}, no. 12, 124034 (2013)
  [arXiv:1307.7977].

\bibitem{Astashenok:2013vza}
  A.~V.~Astashenok, S.~Capozziello, and S.~D.~Odintsov,
  JCAP {\bf 1312}, 040 (2013)
  [arXiv:1309.1978 [gr-qc]].

\bibitem{Mest}
L.~Mestel, {\it Stellar Magnetism} (Oxford University Press, New York, 2012).

\bibitem{Bocquet:1995je}
  M.~Bocquet, S.~Bonazzola, E.~Gourgoulhon, and J.~Novak,
  Astron.\ Astrophys.\  {\bf 301}, 757 (1995)
  [gr-qc/9503044].


\bibitem{magn_stars}
L.~Rezzolla, B.~J.~Ahmedov, and J.~C.~Miller,
  Mon.\ Not.\ Roy.\ Astron.\ Soc.\  {\bf 322}, 723 (2001)
  [astro-ph/0011316];
Y. Tomimura and Y. Eriguchi,  Mon.\ Not.\ R.\ Astron.\ Soc.\  {\bf 359}, 1117 (2005);
  A.~K.~Harding and D.~Lai,
  Rep.\ Prog.\ Phys.\  {\bf 69}, 2631 (2006)
  [astro-ph/0606674];
S.~K.~Lander and D.~I.~Jones, Mon.\ Not.\ R.\ Astron.\ Soc.\  {\bf 395}, 2162 (2009);
P.~D.~Lasky, B.~Zink, K.~D.~Kokkotas, and K.~Glampedakis,
  Astrophys.\ J.\  {\bf 735}, L20 (2011)
  [arXiv:1105.1895 [astro-ph.SR]];
K.~Glampedakis, N.~Andersson, and S.~K.~Lander,
  Mon.\ Not.\ R.\ Astron.\ Soc.\  {\bf 420}, 1263 (2012)
  [arXiv:1106.6330 [astro-ph.SR]];
R.~Ciolfi and L.~Rezzolla,
  Astrophys.\ J.\  {\bf 760}, 1 (2012)
  [arXiv:1206.6604 [astro-ph.SR]].

\bibitem{Hakimov:2013zoa}
  A.~Hakimov, A.~Abdujabbarov, and B.~Ahmedov,
  Phys.\ Rev.\ D {\bf 88}, no. 2, 024008 (2013).


\bibitem{Cheoun:2013tsa}
  M.~K.~Cheoun, C.~Deliduman, C.~Güngör, V.~Keles, C.~Y.~Ryu, T.~Kajino, and G.~J.~Mathews,
  JCAP {\bf 1310}, 021 (2013)
  [arXiv:1304.1871 [astro-ph.HE]].

\bibitem{Astashenok:2014gda}
  A.~V.~Astashenok, S.~Capozziello, and S.~D.~Odintsov,
  Astrophys.\ Space Sci.\  {\bf 355}, no. 2, 333 (2015)
  [arXiv:1405.6663 [gr-qc]].

\bibitem{Konno:1999zv}
  K.~Konno, T.~Obata, and Y.~Kojima,
  Astron.\ Astrophys.\  {\bf 352}, 211 (1999)
  [gr-qc/9910038].

\bibitem{Sotani:2006at}
  H.~Sotani, K.~D.~Kokkotas, and N.~Stergioulas,
  Mon.\ Not.\ R.\ Astron.\ Soc.\  {\bf 375}, 261 (2007)
  [astro-ph/0608626].

\bibitem{DeDeo:2007yn}
  S.~DeDeo and D.~Psaltis,
  Phys.\ Rev.\ D {\bf 78}, 064013 (2008)
  [arXiv:0712.3939 [astro-ph]].

\bibitem{Regge:1957}
T.~Regge and J.~A.~Wheeler,
  Phys.\ Rev.\ {\bf 108}, 1063 (1957).

\bibitem{Haensel:2004nu}
  P.~Haensel and A.~Y.~Potekhin,
  Astron.\ Astrophys.\  {\bf 428}, 191 (2004)
  [astro-ph/0408324].

\bibitem{Naf:2010zy}
  J.~Naf and P.~Jetzer,
  Phys.\ Rev.\ D {\bf 81}, 104003 (2010)
  [arXiv:1004.2014 [gr-qc]].

\bibitem{Ozel:2010fw}
  F.~Ozel, G.~Baym, and T.~Guver,
  Phys.\ Rev.\ D {\bf 82}, 101301 (2010)
  [arXiv:1002.3153 [astro-ph.HE]].

\bibitem{Kiuchi:2007pa}
  K.~Kiuchi and K.~Kotake,
  Mon.\ Not.\ R.\ Astron.\ Soc.\  {\bf 385}, 1327 (2008)
  [arXiv:0708.3597 [astro-ph]].

\bibitem{Potekhin:2013qqa}
  A.~Y.~Potekhin, A.~F.~Fantina, N.~Chamel, J.~M.~Pearson and S.~Goriely,
  Astron.\ Astrophys.\  {\bf 560}, A48 (2013)
  [arXiv:1310.0049 [astro-ph.SR]].

\bibitem{Folomeev:2015aua}
  V.~Folomeev and V.~Dzhunushaliev,
  Phys.\ Rev.\ D {\bf 91}, no. 4, 044040 (2015)
  [arXiv:1501.06275 [gr-qc]].
\end{thebibliography}
\end{document}